\documentclass[twocolumn,preprintnumbers,amsmath,amssymb]{revtex4}
\makeatother
\usepackage[dvips]{graphicx}
\usepackage{amsmath}
\usepackage{amsbsy}
\usepackage{caption}
\usepackage{color}
\DeclareGraphicsExtensions{.pdf,.eps,.png,.jpg,.mps}
\usepackage{gensymb}

\definecolor{textcolor}{cmyk}{0,0,0,1}
\definecolor{magenta}{rgb}{1,0,1}
\definecolor{green}{rgb}{0,1,0}
\definecolor{red}{rgb}{1,0,0}

\usepackage{xcolor}

\begin{document}

\title{
Bonding of Cobalt Atoms, Dimers and Co$_4$ Clusters to Circumcoronene 
}
\author{ T. Alonso-Lanza,$^{1}$, A. Ma\~{n}anes,$^{2}$, and A. Ayuela$^{1}$ }
\affiliation{
$^1$Centro de F\'{\i}sica de Materiales CFM-MPC CSIC-UPV/EHU, Donostia 
International Physics Center (DIPC), Departamento de F\'{\i}sica de Materiales, Fac. de Qu\'{\i}micas, UPV-EHU, 20018 San Sebasti\'an, Spain
\\
$^2$Departamento de F\'{\i}sica Moderna, Universidad de Cantabria, Santander, Spain
\\
}


\begin{abstract}
We study the interaction of circumcoronene molecule C$_{54}$H$_{18}$ with cobalt in the form of atoms, dimers and clusters of four atoms.
We find that the cobalt atom prefers to be on a hollow site in the edge zone. The cobalt dimer is bonded perpendicularly to another different hollow site in the edge zone. The Co$_{4}$ cluster adopts a tetrahedral shape with a face parallel to circumcoronene, placing each of the three atoms over contiguous hollow sites, starting from the edge.
Cobalt remains bonded to circumcoronene as Co$_{2}$ molecules or Co$_{4}$ clusters, rather than spread as isolated atoms, because the cobalt-cobalt interaction is stronger than the cobalt-carbon one.
However, the interaction with cobalt clusters induce small magnetic moment on circumcoronene. The magnetic moment of cobalt counterpart is reduced when it is bonded to circumcoronene.
There is charge transfer between those systems and its direction depends on the relative position of the cluster with respect to circumcoronene.
\end{abstract}

\maketitle

\section{\label{sec:intro} Introduction}

Polycyclic-aromatic hydrocarbons (PAHs) such as benzene and coronene (C$_{24}$H$_{12}$) are found as by-products in combustion processes \cite{richter2000formation} as well as in the interstellar medium \cite{bohme1992pah}.
PAHs show high stability because of the high delocalization of electrons, which allows these molecules to keep its aromaticity \cite{zhou2006novel}, key to understand their interesting properties. The properties of larger PAH molecules like circumcoronene C$_{54}$H$_{18}$ have also been deeply studied \cite{khanna2010atomistic, philpott2009bonding, aryanpour2014electron, dias2002topological}.
Circumcoronene have  $\pi$ delocalized bonds in the form of seven local sextets and six double bonds, which can be represented in a Clar structure \cite{popov2012chemical}.
Different experimental aspects of circumcoronene have been studied such as interaction with ion beams \cite{forsberg2013ions}, loss of fragments \cite{bauschlicher2014loss},  nuclear magnetic resonance (NMR) spectra \cite{ikalainen2008laser}, and electron currents  when applying a perpendicular magnetic field \cite{soncini2003perimeter}.
As commented above for other PAHs,  circumcoronene is specifically thought to be found in the interstellar medium,  accounting for the IR spectra that is experimentally obtained.   The IR spectra is computed in several studies considering neutral and charged circumronene after adsorbing atoms \cite{bauschlicher2000infrared, galue2012electronic, hudgins2005variations}.  Note that atoms can adsorb to circumcoronene on different sites. 

PAHs doped with late transition metals are also in the quest for ultimate memories in spintronics \cite{xiao2009co}.
For smaller PAHs, like benzene and coronene, previous works have studied cobalt adsorption.
When a cobalt atom bonds benzene, it is situated over the center of the ring, giving rise to a $\eta^{6}$ site \cite{belbruno2005half}.
It has also been studied how few transition metal atoms interact with several benzene molecules \cite{pandey2001electronic}. It is known to adopt two kind of structures, multiple-decker or bowl-like, depending on the transition element. For late transition elements, such as cobalt, bowl-like structures are preferred \cite{kurikawa1999electronic}.
Cobalt atom over benzene is predicted in the hollow site \cite{zhou2006novel}.
The magnetism of the cobalt benzene system is mainly determined by the cobalt atom, which displays 1.1 $\mu_{B}$ with a small negative magnetic moment induced in benzene \cite{zhang2008structural}.
Larger coronene can be produced using laser ablation as in interstellar medium \cite{duncan1999production}, characterized using photoelectron spectroscopy \cite{duncan1999production} and collisions with charged ions \cite{lawicki2011multiple}. 
Furthermore, the interaction between iron and coronene has been studied both theoretically by density functional theory calculations \cite{li2008photoelectron} and experimentally by photoelectron spectroscopy \cite{li2008photoelectron} and mass selected laser photodissociation \cite{buchanan1998metal}.
Cobalt and coronene interaction has been well studied both theoretically by density functional theory and experimentally using photoelectron spectroscopy \cite{kandalam2007ground}.
For the much larger circumcoronene, iron atoms preferred bonding  to hollow sites in the edges \cite{bauschlicher2009fe+, simon2010computed}.
However, cobalt atoms deposited on circumcoronene remains yet to be investigated, ranging from single cobalt atoms to small cobalt clusters.

Here we study the graphene flake immediately larger than coronene molecule that keeps an hexagonal shape, which is called circumcoronene. We deposit atoms, dimers, and a small cobalt cluster of four atoms over circumcoronene. This hybrid system can display interesting electronic and magnetic properties due to the interaction of the 3d orbitals of the cobalt atoms with the 2p orbitals of the C$_{54}$H$_{18}$ molecule. 
The possible induced magnetic moment in the graphene flake due to the presence of cobalt atoms and additional properties due to the edges could be of importance for the development of graphene-based nanodevices in the field of spintronics.

\section{\label{sec:model}Computational details}

\begin{figure}[thpb]
      \centering
\includegraphics[width=0.5\textwidth]{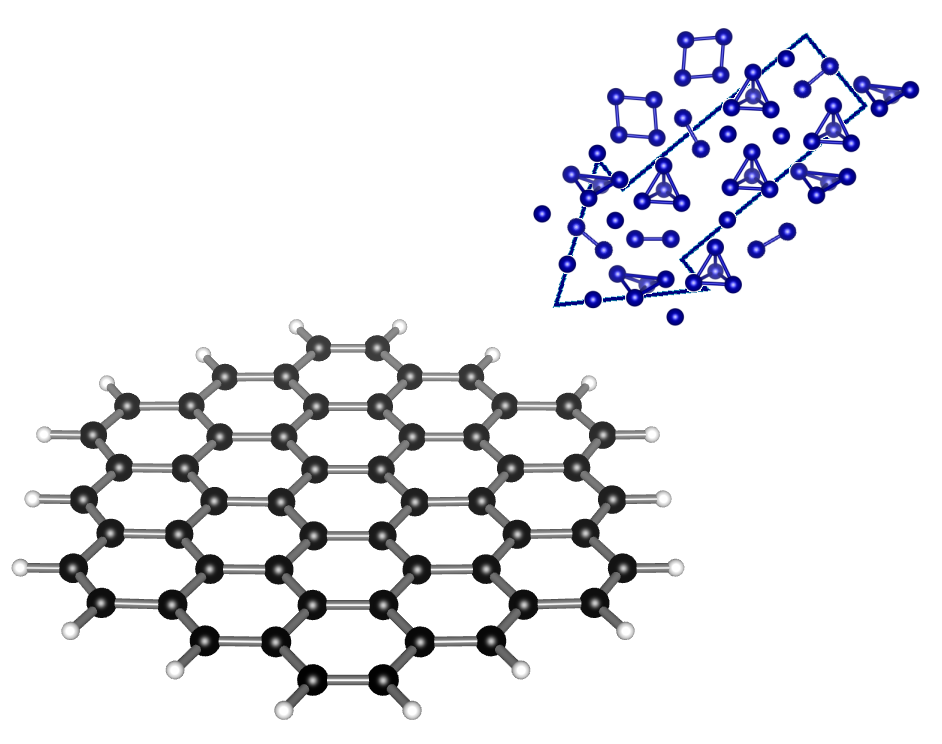}
\caption{\label{experimento} 
Scheme representing the experiment associated to our calculations: cobalt clusters flow towards circumcoronene.} 
\end{figure}

\begin{figure*}[thpb]
      \centering
\includegraphics[width=17cm]{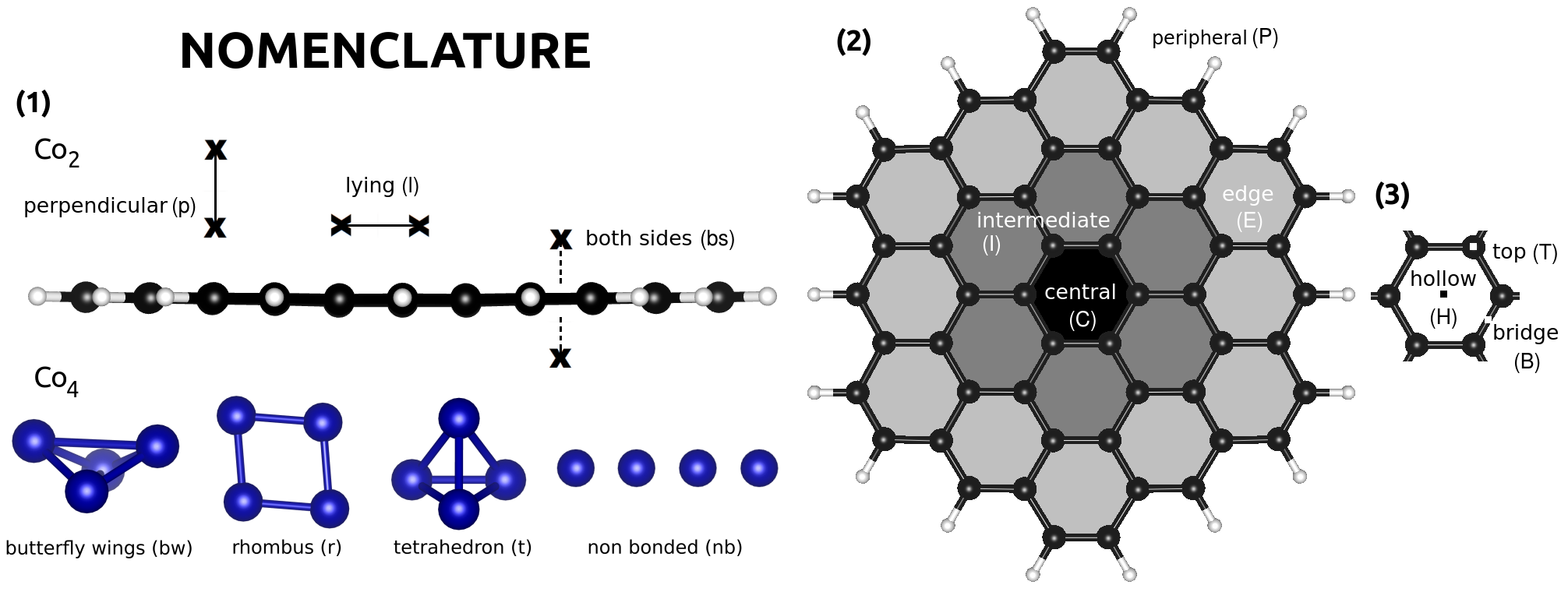}
\caption{\label{nomenclatura} 
Different configurations of Co atoms, Co dimers, and Co$_4$ clusters on circumcoronene. Cobalt atom sites are labelled taking (1) the first character(s) of the name describing the initial disposition of Co$_2$ or Co$_4$ either (p,l,bs) or (bw,r,t,nb), (2) the second label from the circumcoronene region such as (C,I,E,P), and (3) the third character from the hexagonal site. Note that for Co atom there is no option to choose the disposition. When several configurations deserve an equal name, they are distinguished by an order number as commented in text.} 
\end{figure*}

We present here density-functional-theory calculations that solve the Kohn-Sham equations where the electron density is given as the sum of independent particles. We carry out them mainly with the SIESTA method \cite{soler2002siesta}. We choose the generalized gradient approximation for exchange-correlation with the PBE flavor \cite{perdew1996generalized} in order to solve these equations self-consistently. The wavefunctions of independent particles are expanded in basis sets of DZP atomic orbitals. 
We use norm-conserving relativistic Troullier-Martins \cite{troullier1991efficient} pseudopotentials for the core electrons of carbon, hydrogen and cobalt atoms.
The reference electronic configuration for cobalt has 1.50 electrons in the 4s level and 7.50 electrons in the 3d level. We used fractional charges for the occupations of the pseudopotential attending to the electron configurations for the ground state of the free cobalt atom that we obtained in test ADF calculations.  It has been shown that for small transition metal atoms the ground state electronic structure lies in between 3d$^{n-1}$ 4s$^{1}$ and 3d$^{n-2}$ 4s$^{2}$ \cite{moroni1997ultrasoft, johll2009density}.
The electron configurations and the value of the radii were thoroughly tested by computing energy differences between all-electron and pseudopotential calculations for several excited states,, such as 3d$^{7}$ 4s$^{2}$, 3d$^{7}$ 4s$^{1}$ 4p$^{1}$, 3d$^{7}$ 4s$^{1}$ and 3d$^{8}$ 4s$^{1}$.
The pseudopotentials were also tested in real calculations, for example, in cobalt bulks. We obtained an hcp structure, with a first neighbors distance of 2.52 \AA{} and a magnetic moment per atom of 1.65 $\mu_B$, which are in agreement with experimental results and similar all-electron calculations \cite{myers1951spontaneous}.
We tested the pseudopotentials in graphene and benzene molecules obtaining reasonable results for their main properties.
For our isolated systems we choose a large enough unit cell box of 40x40x40 \AA{}, and atoms are fully relaxed. We have tried different cell sizes for the unit cell to ensure that  self-interactions among molecules are negligible. We use the same computational parameters in all the calculations: an electronic temperature of 25 meV and a mesh cutoff of 400 Ry.
Additionally, we used another density functional theory codes to validate our results (Amsterdam Density Functional ADF) \cite{te2001chemistry}, to compute the electron localization function and include the Hubbard term (VASP) \cite{blochl1994projector,VASP,VASP2} and to compute hyperfine parameters (ELK) \cite{dewhurst2016elk}. For the VASP calculations we tried values of  $U_{eff} = U-J$ = 2 and 3 eV, representing a range of reasonable values as found in the literature \cite{piotrowski2011role,vaugier2012hubbard,wang2006oxidation,aykol2014local,jain2011formation} for transition metal systems.
We describe in Fig. \ref{nomenclatura} the nomenclature to specify the different configurations computed for each system with Co atoms, Co dimers, and Co$_4$ clusters.

\section{Results and discussion}

\subsection{Stability}

\subsubsection{Co atom on circumcoronene}
      
\begin{figure*}[thpb]
      \centering
\includegraphics[width=18cm]{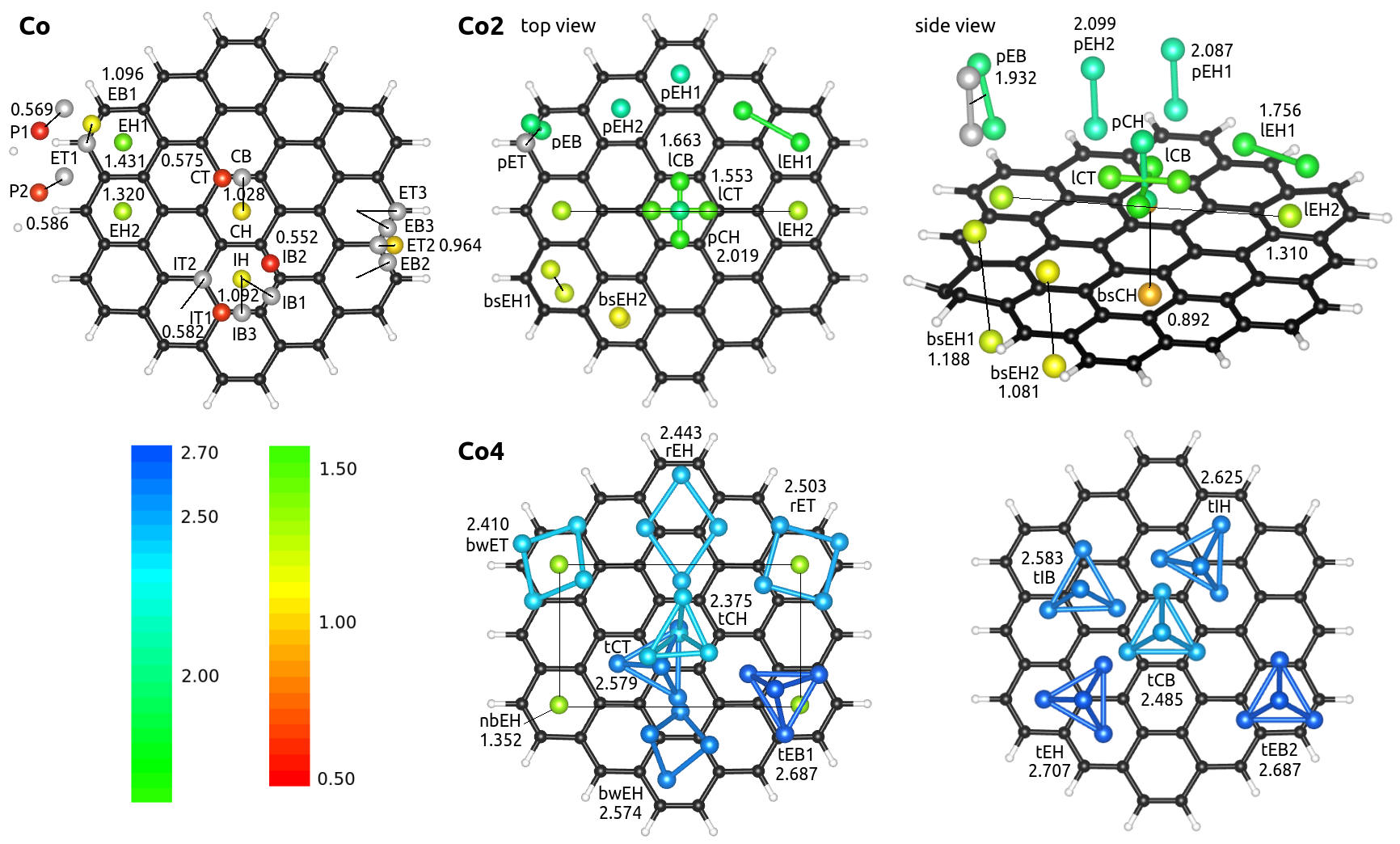}
\caption{\label{geometrias} 
Geometries computed for Co atom, Co$_{2}$ and Co$_{4}$ over C$_{54}$H$_{18}$. Cobalt atoms are colored according to their binding energy E$_b$ in eV, shown close to the name of the geometry. Grey atoms and pEB dimer show initial arrangements that relax to those joined with a thin straight line. For some configurations of Co$_{2}$ and Co$_{4}$, the lines bond Co atoms far away from each other although computed together in special structures (see text).} 
\end{figure*}

We deposit a cobalt atom onto circumcoronene and relax the coordinates starting from several sites. We include top, bridge and hollow sites for the cobalt atom in the three main regions of circumcoronene: central, intermediate and edge. Taking into account that C$_{54}$H$_{18}$ has D$_{6h}$ symmetry, the possible sites reduce substantially. The computed configurations are schematically shown in Fig. \ref{geometrias}. The binding energies per atom are also included. They are calculated as  $E_{b}=(E_{C_{54}H_{18}}+n*E_{Co}-E_{C_{54}H_{18}+Co_{n}})/n$ in eV, where n is the number of cobalt atoms $n$=1,2,4.
Note that several of the input geometries suffer large geometry transformations: some cases relax and fall into other proposed geometries, and few relaxed to even new structures.
It seems clear from the values of the bonding energy that {\it the hollow sites are in general preferred over the top and bridge sites}. For some configurations the cobalt atoms move from the top and bridge sites to the near hollow sites during relaxation, as in the case of CB, IB1, IB3 or IT2 among others. 
Among the hollow sites in the different zones, we find that {\it the farther from the center of circumcoronene the more stable the hollow site becomes}, as given by the sequence CH$<$IH$<$EH2$<$EH1.
In the two peripheral configurations, P1 and P2, the cobalt atom bonds laterally to circumcoronene, in the same plane as carbon atoms. The cobalt atoms are in between the hydrogen and carbon atoms, like catalysts. These geometries are the least stable ones.

Cobalt atoms over hollow sites have a  height about 1.45 \AA{} over circumcoronene. The Co heights increases up to values close to 2 \AA{} for bridge and top sites, like EB1 and IB2. Anyhow, the cobalt-carbon bond lengths for all the sites are about 2 \AA{}. Note that the P1 and P2 configurations, with the cobalt atom almost replacing an hydrogen atom, have shorter Co-C bond lengths of about 1.8 \AA{}, with the cobalt-hydrogen length about 1.5 \AA{}.

We next compare our results with those of literature.
A cobalt atom bonded to benzene on the hollow site is the most stable, having a carbon cobalt distance of 2.02 \AA{} \cite{belbruno2005half}. A hollow site is also predicted for a cobalt atom on graphene and graphene nanoribbons \cite{longo2011ab} using DFT with the generalized gradient approximation \cite{cao2010transition,johll2009density,duffy1998magnetism,mao2008density}. These calculations show heights over the surface of 1.6 and 1.5 \AA{}, and adsorption energies in the range of 0.97-1.34 and 0.97-2.08 eV, respectively. Quantum Monte Carlo method \cite{virgus2012ab} also predicts a hollow site for cobalt atoms in graphene, with a height of 1.49 \AA{}. Using CASSCF to study a cobalt atom over the central hollow site of coronene \cite{rudenko2012adsorption}, a height of 1.6 \AA{} is found. 
A cobalt atom interacting with coronene \cite{kandalam2007ground} finds the most stable configuration in a non-central hollow site, like in our most stable EH1 and EH2 configurations, which are non-central hollow sites.
The stability and sites of all these previously calculated data are in agreement with our results shown above.

\subsubsection{Co$_{2}$ molecule on circumcoronene}

A cobalt dimer bonded perpendicularly to benzene has been proposed as the ultimate magnetic storage device \cite{xiao2009co}. In this section we study Co$_{2}$ dimer  on circumcoronene by considering the geometries shown in Fig. \ref{geometrias}. 
 Between the two possible types of hollow sites around the edges, the side middle (pEH2) is preferred better over the corners (pEH1). Note that the single cobalt atom prefers to be at the corner site. We find that the cobalt dimer binds to PAH systems perpendicularly at hollow sites close to the edges.
The Co heights over the PAH plane for the perpendicular bonds on hollow sites are about 1.6 \AA{} and 3.7 \AA{}. The shortest Co-C distances for each configuration lye within the interval of 2-2.2 \AA{}. The Co-Co distances are about 2-2.1 \AA{}, indicating strong bonds. For the isolated cobalt dimer we obtained approximately 2 \AA{}.
Figure \ref{geometrias} shows that disposing the molecule perpendicular is crucial for stability. Although a hollow site is preferred, the pEB bridge configuration is nearly degenerate. The pEB configuration was initially set perpendicular to circumcoronene, but it ends a little tilted, about 13$\degree$.

Next in stability, we find the configurations lying on circumcoronene. The lEH1 configuration is the most stable among them, and has cobalt atoms over neighbor edge hollows. The distance between cobalt atoms increases up to 2.42 \AA{}, which is approximately 0.3 \AA{} larger than for the perpendicular configurations. The next two most stable configurations place the cobalt molecule parallel to circumcoronene around the central ring, in bridge sites for the lCB case and in top sites for the lCT one, with Co-Co distances of about 2.1 \AA{}.
Comparing these two cases with the pCH configuration, which is also on the central ring, we find hollow, bridge and top sites in a decreasing order of stability. 
The lEH2 case is much less stable with the two cobalt atoms on hollow sites at opposite corners separated by more than 8 \AA{}. 
The next three configurations spread the two cobalt atoms on equivalent hollow sites at each side of circumcoronene, with large Co-Co distances of 3.2 \AA{}. Anyhow, these configurations together with the lEH2 case are being clearly less stable, which means that the cobalt atoms prefer to bond directly between themselves.

A previous work about coronene \cite{kandalam2007ground} found that the most stable configuration have the cobalt dimer perpendicularly on a bridge position at the edge, which could be an analogous to the pEB case, which is indeed highly stable but less than the ground state hollow site. Note that the coronene-cobalt bond length was 2.1 \AA{} \cite{kandalam2007ground}, which is also similar to 2.06 \AA{} of the pEB case.
In agreement with our data, Co$_2$ is perpendicularly bonding graphene over a hollow site with the closest cobalt atom at a height of 1.72 \AA{} and on top a cobalt atom at approximately 2 \AA{}  \cite{cao2010transition,johll2009density}.

\begin{figure*}[thpb]
      \centering
\includegraphics[width=18cm]{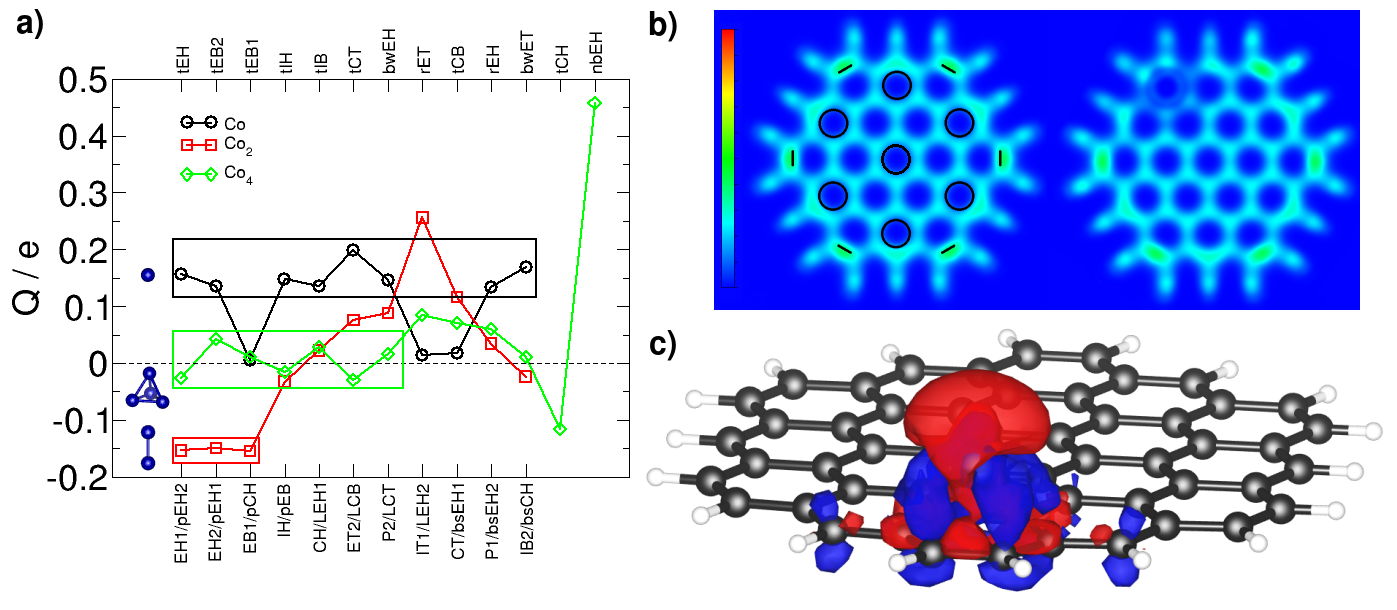}
\caption{\label{cargas} 
(a) Charge transfer Q from cobalt towards circumcoronene for all the computed geometries for Co, Co$_{2}$ indicated in the lower part of the abscissas axis and for Co$_{4}$ indicated in the upper part of the abscissas axis. Positive Q values imply that electrons are gained by circumcoronene. Binding energies decrease from left to right. Colored boxes mark crucial values as commented in the text. (b) Electron localization function (ELF) of circumcoronene (left) and with Co  in the EH1 configuration (right) in a plane approximately 1.2A over circumcoronene. For circumcoronene, sextets and double bonds of the Clar configuration are also plotted. (c) Charge difference in the cobalt atom bonding (EH1 configuration). Blue (red) zones denote zones gaining (losing) charge.} 
\end{figure*}

\subsubsection{Co$_{4}$ molecule on circumcoronene}

In this section we study the stability of Co$_{4}$ clusters interacting with circumcoronene. We consider the three most stable geometries of Co$_{4}$ free clusters, such as butterfly(bw), rhombus(r) and tetrahedron(t), deposited on a corner ring of circumcoronene in the configurations bwET, rET, and tEB1, because the results for Co and Co$_{2}$ on circumcoronene indicate that on edges they would be more stable. The bw and rhombus structures are commensurate on the top and bridge positions of carbon atoms. The tetrahedron becomes the most stable on circumcoronene.

We searched for the most stable tetrahedron arrangement. The tEH configuration has three cobalt atoms over three different hollows and another cobalt atoms on top of them. It becomes the most stable structure.
When the Co$_{4}$ cluster is adsorbed over the center, circumcoronene remains planar. However, when the Co$_{4}$ cluster is placed in the edge zone, circumcoronene bend around the cluster.
We find some general trends after analyzing all the computed structures. First, moving the $Co_4$ cluster from the center to the edges gradually increases the stability. Second, hollow sites are preferred over bridge sites and top sites, which are not even stable in some cases (tCT relax to the hollows).
Third, when computing the nbEH geometry we checked that the cobalt atoms prefer to be clustered rather than to spread over circumcoronene separately. All those trends agree with our results for Co and Co$_{2}$.

The preference for the tetrahedron with one face lying over circumcoronene (tEH) implies large interaction between cobalt and carbon atoms. In the dimer the perpendicular orientation was preferred leaving the farthest cobalt atom almost isolated from carbon atoms. Although it is true that the tetrahedron also isolates one cobalt atom, one can wonder why isolating three cobalt atoms is not preferred (tCH is 0.3 eV higher in energy than tEH).
Here there are two trends to be taken into account. First, a cobalt atom over a hollow site is highly stable as seen for EH1. It seems that the more cobalt atoms over hollow sites the larger the stability. Second, commensurating already bonded cobalt atoms over neighbor hollows imply enlargement of Co-Co bond length, which reduces stability. 
We apply these two ideas to understand Co$_{2}$ and Co$_{4}$ cases. For the case of the cobalt dimer, the bond length varies from 2 \AA{} for the free cluster to 2.42 \AA{} for lEH1, which implies an enlargement of 21\%. The reduction of stability due to the deformation is not compensated by the possible extra cobalt atom over a hollow site. On the other hand, for the cobalt tetrahedron the bond length varies from approximately 2.3 \AA{} for the free cluster to 2.5 \AA{} for tEH, which implies an enlargement of 9\%. As the deformation is lower than for Co$_{2}$ and, in this case, there are two extra cobalt atoms over hollow sites, the configuration with more cobalt atoms interacting directly with carbon atoms is preferred for the Co$_{4}$ case, contrary to the Co$_{2}$ case.

On the other hand, Co$_{4}$ results show many geometries within an interval of 0.4 eV, quite close in energy. Once the cluster is deposited on circumcoronene, it could slide easily with three cobalt atoms in a plane parallel to circumcoronene. However, additional calculations in order to compute the barriers between those states are beyond the actual scope of this work.
Note that the interaction between Co$_{4}$ clusters and graphene has been studied previously \cite{johll2011graphene}, obtaining a geometry similar to tEH, in agreement with the ground state obtained here.

\subsection{Charges and bonding}

\subsubsection{Charge transfer}

We study the charge transfer from Co atoms to circumcoronene which is shown in Fig. \ref{cargas}. In the Pauling scale of electronegativity, carbon and cobalt have values of 2.55 and 1.88, which means that charge transfer from cobalt towards carbon should be expected. Indeed, this happens for most of the cases. 
Electrons are gained by circumcoronene when a cobalt atom is placed over it, for instance, 0.16 electrons for the ground state (EH1). Similar values are obtained for other configurations except for top sites, such as the IT1 and CT cases with a charge transfer approximately ten times smaller, and for the EB1 configuration with an almost zero value because the cobalt atom is over a double bond.
Nevertheless, when Co$_{2}$ dimer bonds perpendicularly to a circumcoronene hollow (pEH2,pEH1,pCH), about 0.15 electrons pass to cobalt atoms, mainly to the farther atom. When Co$_{2}$ lies parallel (lEH1,lCB,lCT), circumcoronene slightly gains charge. Note that the lEH2 case presents a large value of 0.256 electrons, as the sum of two EH2 single Co atom configurations.
For Co$_{4}$ the charge transfer is very small in general. Even though, it can be seen that, for tetrahedra, when cobalt atoms are over hollow sites (tEH,tIH,tCT(which relaxed to hollows),tCH) they gain electrons, while when they are over bridge sites (tEB2,tEB1,tIB,tCB) circumcoronene gains electrons. A relatively large value is found for the tCH case with only one cobalt atom bonding a hollow site. As for the lEH2 case, which behaves like two EH1 configurations, the nbEH configuration behaves like four EH1 atom configurations concerning the amount and the sign of the charge transfer.
It does not seem to exist a clear correlation between the amount of charge transferred and the stability within each configuration of Co atoms, dimers and clusters.

\subsubsection{Bonding}

We comment briefly on ideas related the bonding bringing into contact the local charges.
We firstly study the preference of cobalt for the circumcoronene edges. 
We start by considering the aromaticity of the molecule. The well-known Huckel's rule applies to circumcoronene. Each of the 54 carbon atoms contribute with one electron from the 2p$_{z}$ orbital, so this rule estimates that a planar ring molecule like circumcoronene has aromatic properties when having $4n+2$ delocalized electrons with n=13. Following Clar's rule, the resonance structure of circumcoronene, shown by circles representing aromatic bonds and lines denoting double bonds in Fig. \ref{cargas}b (left part), has the largest number of disjoint aromatic $\pi$ sextets with double bonds found in the edges. However, the electron localization function in Fig. \ref{cargas}b (left part) shows that, although the double bonds are present, there is also charge accumulation at other carbon atoms in the edge, meaning that the Clar structure is not fully reproduced. As a result, circumcoronene has charge accumulation all over the carbon atoms in the edges, explaining the preference of cobalt atoms to be located there.
A related argument for the preference for being at the edges would be the loss of symmetry. For the central hollow configuration of the cobalt atom (CH) the 3d$_{xz}$ and the 3d$_{xz}$ levels of cobalt are degenerated, presenting a minority spin level at the Fermi energy. When the cobalt atom is over an edge hollow (the EH1 and EH2 configurations), those levels become different and shift away from Fermi energy to larger energies, and the system gains stability.

Secondly, we comment on the preference for hollow over bridge and top sites. There are two main differences between hollow and top sites: hybridization of cobalt 3d levels with p$_{z}$ states from carbon atoms and filling of cobalt 4s levels. As we show in the next section, the total magnetic moment increases 2 $\mu_{B}$ from hollow to top, with some contribution from 4s levels and more than 1 $\mu_{B}$ from 3d levels, which suffers a large rehybridization when moving from hollow to top and vice versa. 
The top site electronic structure is much more similar to that of a free cobalt atom than the hollow one. The large charge reorganization has been argued as the reason for the largest stability for the hollow instead of the top site \cite{johll2009density}.
Similarly, the preference for a perpendicular bonding of Co$_{2}$ is explained due to the larger charge change suffered by the farthest cobalt atom; when the dimer is parallel to circumcoronene both cobalt atoms remain similar to free state. The extra energy needed to recover the free state configuration explain the stability of hollow and perpendicular disposition.

We now focus on the charge reorganization for the most stable cobalt atom over circumcoronene (EH1). The ELF in Fig. \ref{cargas}b (right part) shows that charge from the bonds between carbon atoms is removed. This is in good agreement with Fig. \ref{cargas}c, where the red lobes denote weakening of the bonds between the carbon atoms. Furthermore, there is a charge transfer from the upper part of the cobalt atom, denoted by a big red lobe, to the areas which represent the six cobalt carbon bonds, as blue lobes show.

\subsection{Magnetism}

\subsubsection{General remarks}

\begin{figure*}[thpb]
      \centering
\includegraphics[width=18cm]{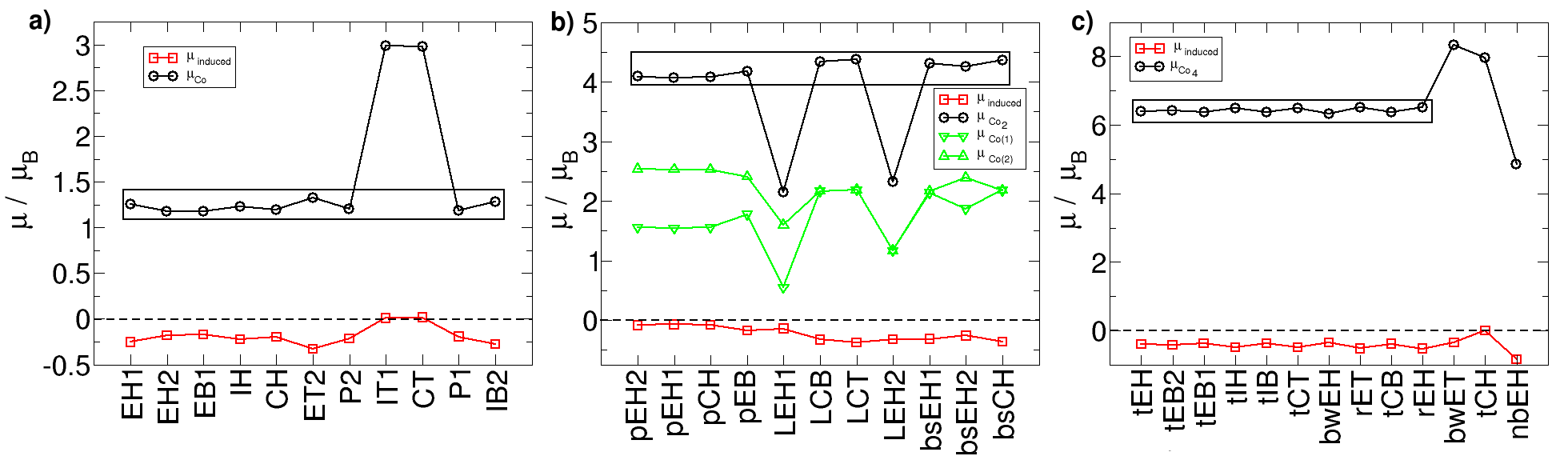}
\caption{\label{magnetism} 
Magnetic moments of cobalt and induced magnetic moments in circumcoronene for all the computed configurations for (a) Co, (b) Co$_{2}$, and (c) Co$_{4}$ on circumcoronene. For the Co$_{2}$ cases, Co(1) (Co(2)) refers to the cobalt atom closer (farther) to circumcoronene. Binding energies decrease from left to right. Black rectangles mark representative values commented in the text.} 
\end{figure*}

We compute the magnetic moment on each atom as the difference between the Mulliken charge up and the Mulliken charge down. We present in Fig. \ref{magnetism} the magnetic moments of the cobalt atoms and cluster and the induced magnetic moment on circumcoronene. In most of the cases the total magnetic moment is 1 $\mu_{B}$ when a cobalt atoms bonds circumcoronene, 4 $\mu_{B}$ when it is a cobalt dimer and 6 $\mu_{B}$ for the tetramer. We have checked that the net magnetic moment is well preserved when adding the Hubbard term U to Co atoms for the most stable geometry of each case. Among cobalt atoms, those farther to circumcoronene show larger local magnetic moments (2.5 for Co$_{2}$) because the closer cobalt atoms shield the effect of carbon atoms. For almost all the cases there is negative induced magnetic moment on circumcoronene: -0.25 $\mu_{B}$ for EH1, almost zero for the dimer bonding circumcoronene perpendicularly and -0.4 $\mu_{B}$ for the tEH1 configuration.

Our results mainly agree with literature. For a cobalt atom over a graphene hollow \cite{cao2010transition,johll2009density,duffy1998magnetism,mao2008density} the cobalt magnetic moment is 1.15 $\mu_{B}$. Using quantum Monte Carlo \cite{virgus2012ab} the total magnetic moment is 1 $\mu_{B}$. It is also 1 $\mu_{B}$ when studying a cobalt atom over coronene \cite{rudenko2012adsorption,kandalam2007ground} using CASSCF.
The total magnetic moment is reported to be 4 $\mu_{B}$ for Co$_{2}$ over coronene \cite{kandalam2007ground} and graphene \cite{cao2010transition,johll2009density}, with 1.58 and 1.66 $\mu_{B}$ on the close cobalt atom to graphene respectively.
For a tetrahedron similar to tEH over graphene \cite{johll2011graphene} a total magnetic moment of 6 $\mu_{B}$ is reported.

Now we comment briefly on the special cases. For a cobalt atom over the IT1 and CT top sites the total magnetic moment is increased to 3 $\mu_{B}$ due to the population of the 4s up level. For Co$_{2}$ it is 2 $\mu_{B}$ for lEH1 and lEH2 and for Co$_{4}$ it is 4 $\mu_{B}$ for nbEH. In these three cases the cobalt atoms are largely separated between them, and the value of the magnetic moment can be explained as the sum of two and four individual atoms respectively. For Co$_{4}$ the bwET and tCH cases display larger magnetic moment (8 $\mu_{B}$) because they leave more cobalt atoms farther from circumcoronene.

\subsubsection{Experimental proposal to measure hyperfine interactions}

\begin{figure*}[thpb]
      \centering
\includegraphics[width=18cm]{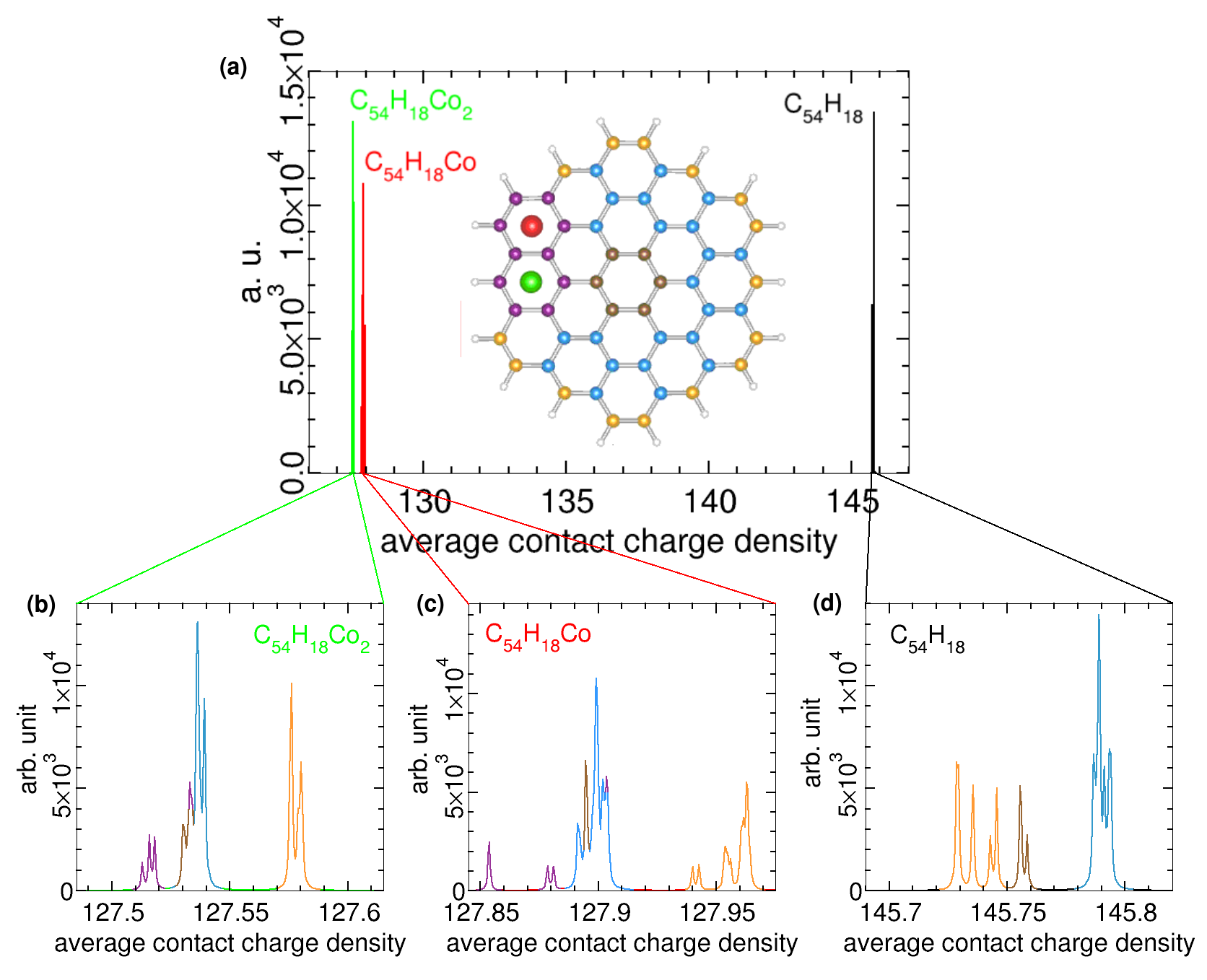}
\caption{\label{hyperfine} 
(a) Relative intensity of the average charge contact density expressed in atomic units for ground state configurations of C$_{54}$H$_{18}$Co$_{2}$, C$_{54}$H$_{18}$Co and C$_{54}$H$_{18}$, which are enlarged in (b), (c) and (d), respectively. In order to gain further insight, panel (a) displays colored atoms attending to the zones edge, interior, central and ring where Co atom is bonded.} 
\end{figure*}

Electric and magnetic fields internal to the atom interact with the nucleus. This interaction produces as a consequence the hyperfine structure. We compute here hyperfine parameters that could be looked at in experiments. There are two different terms that constitute the hyperfine interaction: quadrupole splitting and magnetic splitting. Apart from those, the spectra is shifted respect to its center of gravity due to the so-called M{\"o}ssbauer isomeric shift. This shift depends on the chemical environment of the nuclei and it measures the interaction of the \textit{s} electrons with the nucleus because they have non zero probability of being in the nucleus. The isomer shifts are determined by an average of the charge density within the nuclear volume \cite{shenoy1978mössbauer}.
To measure this interaction, we use the ELK code to compute the hyperfine parameters\cite{blugel1987hyperfine}. Among them we find the average contact charge density value\cite{blugel1987hyperfine}, which measures the charge density in a close region to the nucleus. We focus on the value of the average contact charge density expressed in atomic units on each carbon atom for three different cases, namely the ground state structures of C$_{54}$H$_{18}$, C$_{54}$H$_{18}$Co and C$_{54}$H$_{18}$Co$_2$.

In order to give some guidelines that could help to differentiate the presence of cobalt bonded to circumcoronene, we fit a lorentzian to the data obtaining the spectrum for each case. The spectra for the three cases are shown together in Fig. \ref{hyperfine} (a), which also contains an inset with a pattern colored correlating the origin of the peaks to their molecule sites. Firstly, let us focus on the spectrum for the bare circumcoronene in Fig. \ref{hyperfine}(d). It is possible to differentiate between the atoms that are in the edge and the rest of them colored as orange and blue peaks, respectively. There is also a main peak representing the six carbon atoms in the central ring.
Within the peaks originated at the edges sites there should be two different peaks which reflects the two different types of atoms. However, they split because the average contact charge density is extremely sensitive to distances and there are tiny differences between some directions. For the blue zone we found four different peaks, which correspond to the four different distances to the center for the blue-colored carbon atoms.

Figure \ref{hyperfine}(a) shows also the two cases in the presence of Co. The spectra suffers a remarkable shift towards smaller values when cobalt binds circumcoronene. Due to the presence of one more cobalt atom the shift for C$_{54}$H$_{18}$Co$_2$ is larger than the one for C$_{54}$H$_{18}$Co. We can look in detail for these cases in Fig. \ref{hyperfine}(b) and (c).
To remark that there is an inversion between the two main peaks (orange and blue) when adding cobalt. The blue peak appears on the left part due to larger shift than the orange one. 
In purple we mark the six carbon atoms that constitute the ring over which the cobalt atom bonded to carbon is located. These peaks are more shifted towards smaller values because of the average contact charge density. Notice that different peaks appear for Co and Co$_2$ due to the different distances to the center.

\section{Summary and conclusions}

We have studied the interaction of cobalt atoms, dimers and small clusters of four atoms with circumcoronene. Some general trends rise such as the preference of cobalt to bond hollow sites located at the edge of the molecule. Further study revealed the charge accumulation in the edges as the cause of this behavior. We found large charge reorganization that produces reduced magnetic moment, and accounts for the preference for hollow sites.
The bonding mechanism over a hollow site consist on weakening carbon carbon bonds with six carbon cobalt bonds. Finally, we propose a possible experiment that could allow to distinguish the presence of cobalt atoms doping circumcoronene based on the isomeric shift attending to the different environment of carbon atoms.

\begin{acknowledgments}

This work has been partially supported by the Projects FIS2013-48286-C02-01-P and FIS2016-76617-P of the Spanish Ministry of Economy and Competitiveness MINECO, the Basque Government under the ELKARTEK project(SUPER), and the University of the Basque Country  (Grant No. IT-756-13). TA-L acknowledge the grant of the MPC Material Physics Center - San Sebasti\'an.

\end{acknowledgments}

\clearpage

\section*{Supplemental Information I: SIESTA calculations: Free molecules and clusters}
\subsection{Circumcoronene (C$_{54}$H$_{18}$)}

\begin{figure}[thpb]
      \centering
\includegraphics[width=8cm]{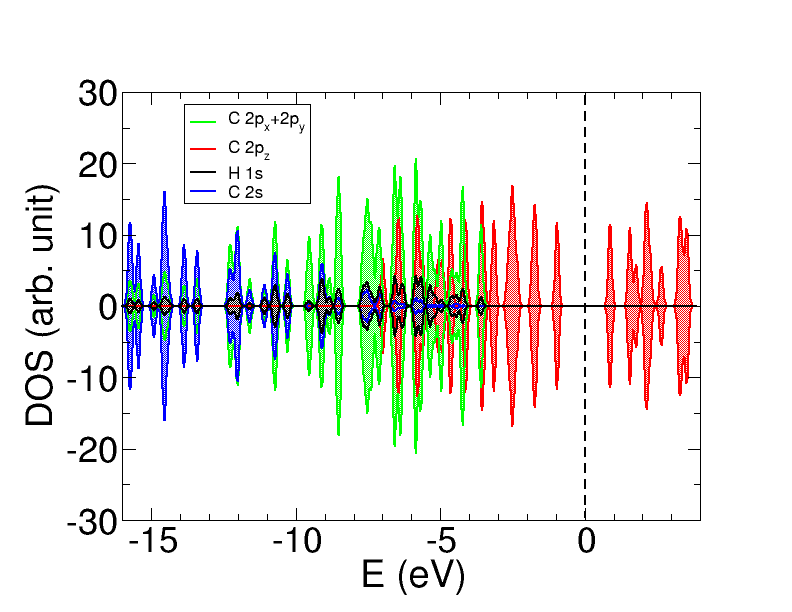}
\caption{\label{fig:circumcoronene} 
Projected density of states for C$_{54}$H$_{18}$ circumcoronene.} 
\end{figure}

Among poly-aromatic hydrocarbons (PAH), the circumcoronene molecule is composed of 54 atoms of carbon and 18 atoms of hydrogen, with lateral dimension slightly smaller than 1.5 nm.
With respect to the well-known coronene C$_{24}$H$_{12}$, the structure of circumcoronene has one more external layer, keeping hexagonal with the carbon dangling bonds saturated by hydrogen atoms in a well-defined D$_{6h}$ symmetry.

Concerning bond lengths in circumcoronene, we found that the position of the C-C bond is related to its length. The bonds around the central point of circumcoronene, taken as reference, have a length of 1.43-1.44 \AA{}, slightly larger than those of pristine graphene (1.42 \AA{} \cite{neto2009electronic}) and  benzene (1.39 \AA{}) \cite{coulson1947energy, pandey2000unique}. 
As we move towards the edge, larger and shorter values are alternating. 
This can be easily understood by the aromatic structure of the molecule. The shortest bond lengths correspond to the double bonds located at the six corners.

Studying the aromaticity of the molecule, the well-known Huckel's rule applies to circumcoronene. Each of the 54 carbon atoms contribute with one electron from the 2p$_{z}$ orbital, so this rule estimates that a planar ring molecule like circumcoronene has aromatic properties when having $4n+2$ delocalized electrons with n=13.
Following Clar's rule, the resonance structure of circumcoronene, shown in Fig.~\ref{fig:circumcoronene}, has the largest number of disjoint aromatic $\pi$ sextets with double bonds found in the edges.

About the carbon hydrogen bonds, there are eighteen bonds about 1.11 \AA{}, a value which agrees with the typical bond length for carbon hydrogen bonds in graphane \cite{sofo2007graphane} and benzene (which is 1.09 \AA{} \cite{pandey2000unique}).

To study the stability of the system we compute the bonding energy $E_{b}$ with respect to the atoms. It is defined as $E_{b}=(54E_{C}+18E_{H}-E_{C_{54}H_{18}})$ where $E_{C}$ is the total energy of a single C atom, $E_{H}$ is the total energy of a single Co atom and $E_{C_{54}H_{18}}$ is the energy of circumcoronene.
We obtain a total bonding energy of 464.982 eV, which is due to 54 carbon atoms and 18 hydrogen atoms, so that per atom it gives 6.45 eV/atom in the order of covalent bonds. This means that the molecule object of study is highly stable.
Although the gap is well-known to be underestimated within the exchange- correlation potential used within GGA-DFT, it is about 1.84 eV for the circumcoronene molecule, still far from the zero gap of graphene. 

There is no spin polarizations, as seen in Fig~\ref{fig:circumcoronene},  because the molecule edges does not have spin.  
We have computed circumcoronene (both with SIESTA and ADF) setting an initial state with locally polarized states on the edges in a way that has been suggested previously \cite{fernandez2007magnetism}, and the final state is locally unpolarized, in agreement with Fern{\'a}ndez-Rossier et al \cite{fernandez2007magnetism} predictions.
The circumcoronene isolated molecule has zero-net magnetic moment. This agrees with Lieb's theorem, which for circumcoronene would predict zero spin polarization due to the fact that there is the same number of atoms on each sublattice \cite{fernandez2007magnetism}. Nevertheless, it is possible for a system to have partial magnetic moment on their atoms and to have a value of zero for the total magnetic moment of the system.
Experimentally it is possible to obtain the spatial distribution of the edge states spins in the graphene nanosheets using the electron spin resonance (ESR) technique at a temperature lower than 20 K \cite{joly2010effect}. The shape of the nanosheet determine both the local magnetic moments and the net magnetic moment. In this type of molecules, the magnetic moment would be localized in the edges, and it would be due to spin-polarized non-bonding states. When the edges are magnetic, there is a strong ferromagnetic (FM) coupling between the edge states that belong to the same sublattice and a weaker FM or antiferromagnetic (AFM) coupling between those of different sublattice. The atoms in a zigzag edge belong to the same sublattice and so they have ferromagnetic coupling. If we apply it to circumcoronene, all the atoms of each side in a hexagonal shape belong to the same sublattice, and the atoms in the neighbor side to another sublattice. Thus, if we found local magnetism in the edge of circumcoronene we could expect to have each side alternatively polarized.
However, the existence of local magnetic moments depends not only on the size of the nanosheet but also in the shape.
Kumazaki et al \cite{kumazaki2008local} used a Hubbard model to study magnetism on the edges, obtaining that vacancy states in the same sublattice can generate local magnetic moments; they proposed that three consecutive vacancies of the same sublattice are enough to obtain local magnetic moment.
Attending to this, circumcoronene should present local magnetic moment.
However, Fern{\'a}ndez-Rossier et al \cite{fernandez2007magnetism} proposed a minimum size of eight atoms per side to find locally polarized states \cite{fernandez2007magnetism}.
Since circumcoronene has three edge atoms along one edge, it is expected to find still locally unpolarized states.

We analyze Fig~\ref{fig:circumcoronene} in detail looking at the projected density-of-states for the s states for the carbon and hydrogen atoms and for the p states for the carbon atoms. The s states from hydrogen and carbon are found at the same low energies because they are hybridized. We differentiate between the 2p states that are in the plane (2p$_{x}$ and 2p$_{y}$) and the 2p states that are perpendicular to circumcoronene (2p$_{z}$).
The p-type states lying on the plane of circumcoronene are also hybridized with the s-states, in the so called sp$^{2}$ hybridization, leaving the 2p$_{z}$ non-hybridized.
The 2p$_{z}$, located at higher energies, are both occupied and empty states around the Fermi level, so they would be held responsible for bonding cobalt atoms.

\subsection{Free cobalt clusters}

{\it 
Co atom.} We compute the cobalt atom using the SIESTA code and we obtain a magnetic moment of 3 $\mu_{B}$. We find that the 4s and 5d levels have lower energy when they are occupied by 1.6 and 7.4 electrons, respectively. We use this configuration to generate Co pseudopotential.

{\it 
Co$_{2}$ molecule. }
Using the generated cobalt pseudopotential, we compute the cobalt dimer. We obtain a relaxed bond length of 2.036 \AA{}, a bonding energy of 1.139 eV, and a magnetic moment about 5 $\mu_{B}$. Johll et al \cite{johll2009density} report a bond length of 1.94 \AA{} and a total magnetic moment of 4 $\mu_{B}$ for the cobalt dimer performing DFT calculations using ultrasoft pseudopotentials. Seifert et al \cite{xiao2009co} obtain also a value of 4 $\mu_{B}$ for the magnetic moment of the cobalt dimer, using an all-electron full-potential local-orbital scheme (FLPO) code. But the magnetic moment of the cobalt dimer has also been reported to be 5 $\mu_{B}$ when performing DFT calculations using the BLYP functional \cite{zhang2008structural}. See next Appendix searching on the comparison with our ADF test results.

{\it 
Co$_{4}$ cluster}

\begin{figure}[thpb]
      \centering
\includegraphics[width=8cm]{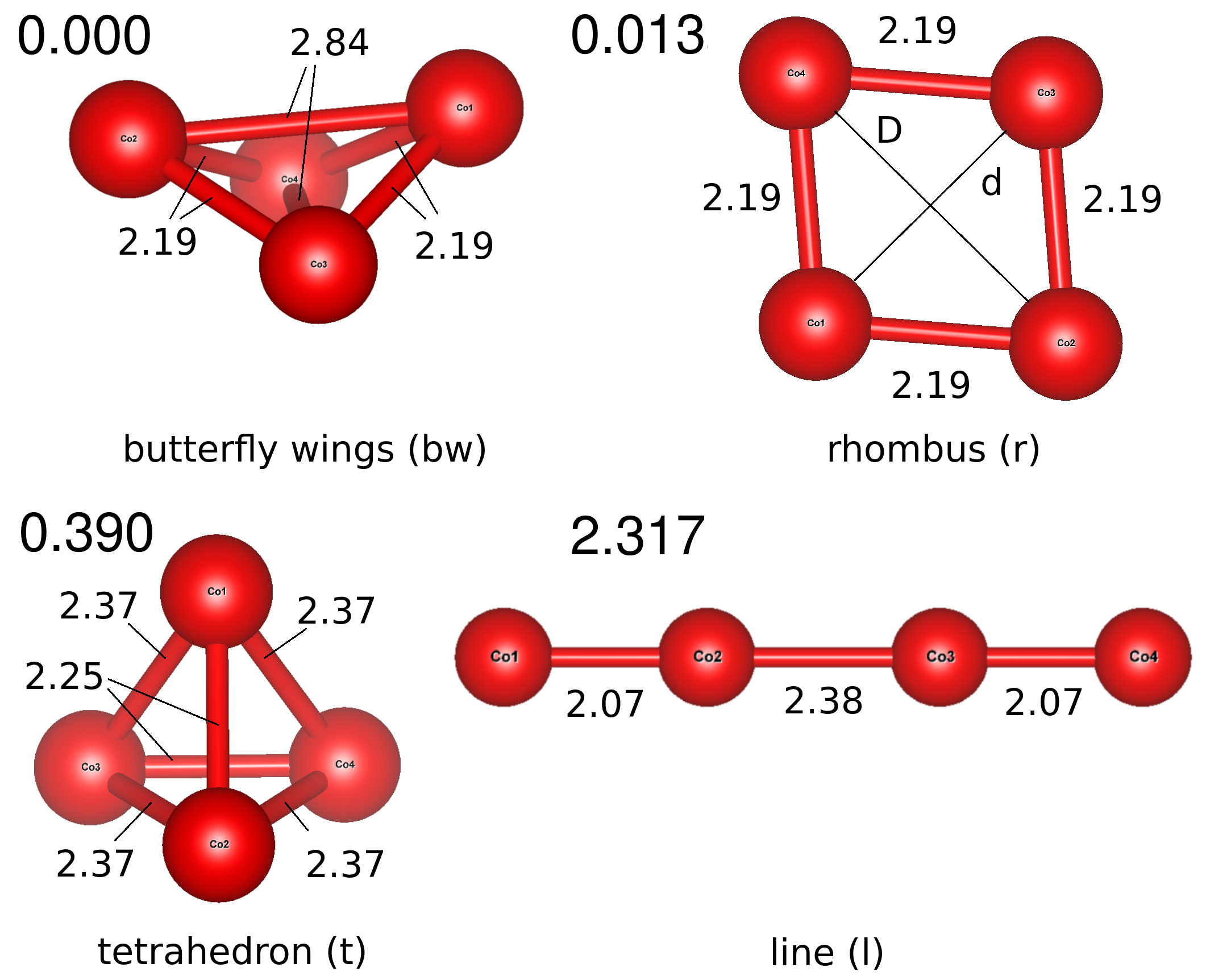}
\caption{\label{tetramer} 
Stability order for free Co$_{4}$ isomers. The values of the relative energy respect to the ground state are shown for each isomer. For the rhombus isomer, the values of the diagonals are D = 3.34 \AA{} and d = 2.85 \AA{}.}
\end{figure}

We check four different geometries for the free Co$_{4}$ cluster in order to find the low lying isomers with minimum energy: two different tetrahedra, a square and a line, shown in Fig. \ref{tetramer}.
The regular tetrahedra has initial edges 2.25 \AA{} long, following the equilibrium geometry in Ref. \cite{yoshida1995spin}, where a distorted tetrahedron had three edges of 2.25 \AA{} and the other three with 2.37 \AA{}.
The input regular square has sides of 2.1 \AA{}, and in the linear structure  all the cobalt atoms are separated by 2.1 \AA{}.
The most stable structure is a non-regular tetrahedron that we called butterfly wings (BW) because of its shape. Note that it is obtained by fully relaxing the input regular tetrahedron, which suffers a very large structural transformation. It has a bonding energy per atom of 2.078 eV. The second most stable isomer is the square with a close bonding energy of 2.075 eV, nearly degenerated, which ends in a rhombus like structure.
The third most stable isomer is a less distorted tetrahedron with a bonding energy of 1.980 eV. All these isomers are largely preferred over the linear structure, which presents quite lower bonding energy, 1.499 eV.
Since the linear structure is much less stable, it means that maximizing the number cobalt-cobalt bonds increases stability. However, in the most stable structure after relaxation there are two kinds of bonds: ones that becomes much larger and other much shorter \cite{datta2007structure}, the latter allow for large 3d bonding interaction. This effect looks like Jahn-Teller distortion.

The total magnetic for tetrahedra isomers is about 10 $\mu_{B}$ except for the less distorted tetrahedra, which has a slightly larger value of almost 11 $\mu_{B}$.
The four atoms are equivalent from the magnetic point of view, except the BW isomer with local moment values of 2.500 $\mu_{B}$ and 2.672 $\mu_{B}$.
For the rhombus structure, the atoms opposed by the large diagonal have a magnetic moment of 2.43 $\mu_{B}$, while the atoms opposed by the shorter diagonal have close values of 2.58 $\mu_{B}$
For the linear structure, the atoms at the ends have a larger value of the magnetic moment (2.626 $\mu_{B}$) than the atoms in the center (2.374 $\mu_{B}$).

All these findings are in agreement with a large amount of publications on transition metal clusters and, specifically, on cobalt clusters.
Focusing on Co$_{4}$ cluster,  the most stable energy structure is a
tetrahedron \cite{yoshida1995spin, datta2007structure, rodriguez2003structure,
  johll2011graphene, andriotis1998tight} as we obtain in our
calculations. However, this tetrahedron structure is reported to have
differences. For example, in Ref. \cite{rodriguez2003structure} the
tetrahedron cluster remains non-distorted with a magnetic moment per atom of 2.99 $\mu_{B}$, and an average bond distance of 2.34 \AA{}. Andriotis et al \cite{andriotis1998tight} obtained a tetrahedron with bond lengths between 2.61-2-80 and a magnetic moment of 2.50 on each cobalt atom. In Ref. \cite{datta2007structure} they also obtained that the most stable is a highly distorted tetrahedron followed by a rhombus, the two structures being much more stable than the linear one. Concerning the magnetic moment \cite{datta2007structure}, all of the cobalt atoms in the tetrahedron and the rhombus have the same value, 2.5 $\mu_{B}$.
Johll et al \cite{johll2011graphene} proposed a structure with two bond lengths of 2.14 and 2.74 \AA{}, in agreement with the one we found here, and also the total magnetic moment (10 $\mu_{B}$).

\section*{Supplemental Information II: Full electron ADF results brought into contact with SIESTA ones}

In this section we begin by showing test results obtained using the ADF code. We used a frozen core treatment of the inner shell.
The equations are solved using localized basis set of Slater type orbitals for the wave functions and the generalized gradient approximation for the exchange and correlation energy, using the revPBE expression. For circumcoronene with Co atoms, Co dimers, and Co$_4$ clusters we test some of the configurations with cobalt around the central hollow. These geometries are the most appropriate to compare results with those of graphene because the central hollow is the farthest position from the edges, and in some way it can be expected a compensation of the border effects due to the symmetry.
Obtained with a code that uses all electron basis sets, these calculations serve us next to check the results from the SIESTA code calculations in order to assess the validity of the pseudopotentials employed.
We collect the ADF results for these center configurations in Table~\ref{tbl:adf}.

\begin{table*}
\begin{tabular}{|c|c|c|c|c|c|c|c|c|c|c|c|c|}
\hline
\multicolumn{13}{|c|}{$C_{54}H_{18} + Co$}\\
\hline
 conf & $E_{b}$/eV   & \multicolumn{2}{|c|}{$z_{Co_{1}}$/\AA{}} & \multicolumn{2}{|c|}{$d_{Co-C}$/\AA{}} & \multicolumn{2}{|c|}{$\mu_{T}$/$\mu_{B}$}   & \multicolumn{2}{|c|}{$\mu_{Co}$/$\mu_{B}$} & \multicolumn{2}{|c|}{$\mu_{ind}$/$\mu_{B}$}  & Q/e$^{-}$ \\
\hline  
 CH  & 1.000 & \multicolumn{2}{|c|}{1.516}   & \multicolumn{2}{|c|}{2.138}    & \multicolumn{2}{|c|}{1.000} &  \multicolumn{2}{|c|}{1.151}  & \multicolumn{2}{|c|}{-0.151}   & 0.11\\
\hline
\multicolumn{13}{|c|}{$C_{54}H_{18} + Co_{2}$}\\
\hline  
 conf & $E_{b}$/eV   & $z_{Co_{1}}$/\AA{} & $z_{Co_{2}}$/\AA{} & $d_{Co-C}$/\AA{} & $d_{Co-Co}$/\AA{} & $\theta$/$ ^{0}$ &  $\mu_{T}$/$\mu_{B}$ & $\mu_{Co}$/$\mu_{B}$ & $\mu_{Co_{1}}$/$\mu_{B}$ & $\mu_{Co_{2}}$/$\mu_{B}$ & $\mu_{ind}$/$\mu_{B}$  & Q/e$^{-}$ \\

\hline
 pCH  & 1.779 & 1.794 & 4.161  &  2.281 & 2.368 & 0.00 & 0.000 & -0.078 & -2.007 & 1.929 & 0.078 & -0.216\\
 lCB  & 1.700  & 2.289 &  2.289 &  2.141 & 2.168 & 90.00 & 4.000 & 4.308 & 2.154 & 2.154 & -0.308 & 0.083\\
 lCT   & 1.598 & 2.291 & 2.291 & 2.137  & 2.191 & 90.00 & 4.000 & 4.328 & 2.164 & 2.164 & -0.328 & 0.062\\
\hline
\multicolumn{13}{|c|}{$C_{54}H_{18} + Co_{4}$}\\
\hline
 conf & $E_{b}$/eV   &   \multicolumn{4}{|c|}{$d_{Co-C}$/\AA{} / $\mu_{Co}$/$\mu_{B}$ Co(1)-Co(2)-Co(3)-Co(4)}   & \multicolumn{2}{|c|}{$\mu_{T}$/$\mu_{B}$}   & \multicolumn{2}{|c|}{$\mu_{Co}$/$\mu_{B}$} & \multicolumn{2}{|c|}{$\mu_{ind}$/$\mu_{B}$}  & Q/e$^{-}$ \\
\hline  
 tCB  & 2.547 &  \multicolumn{4}{|c|}{2.112/1.475 - 2.112/1.475 - 2.112/1.475 - 3.989/1.943}   & \multicolumn{2}{|c|}{6.000}    & \multicolumn{2}{|c|}{6.369} &  \multicolumn{2}{|c|}{-0.369}     & 0.154\\
 bwCB & 2.442  & \multicolumn{4}{|c|}{2.254/1.700 - 2.254/1.700 - 3.507/2.403 - 3.507/2.403}   & \multicolumn{2}{|c|}{8.000}    & \multicolumn{2}{|c|}{8.206} &  \multicolumn{2}{|c|}{-0.206}   & 0.015\\
 bwCT & 2.425  & \multicolumn{4}{|c|}{2.258/1.748 - 2.258/1.748 - 3.650/2.384 - 3.650/2.384}   & \multicolumn{2}{|c|}{8.000}    & \multicolumn{2}{|c|}{8.264} &  \multicolumn{2}{|c|}{-0.264}   & 0.059\\
 \hline
\end{tabular}
\caption{\label{tbl:adf}
ADF results for all the computed configurations. Values of the bonding energy $E_{b}$, the height of the cobalt atom respect to circumcoronene $z_{Co}$, the distance between the cobalt atom and its first carbon neighbor $d_{Co-C}$, the total magnetic moment $\mu_{T}$, the magnetic moment of the cobalt atom $\mu_{Co}$, the induced magnetic moment in circumcoronene $\mu_{ind}$ and the amount of electrons gained by circumcoronene Q (positive (negative) values mean electrons gained (lost) by circumcoronene).
For Co$_{2}$ clusters the next values are also shown: the height of the cobalt atoms respect to circumcoronene $z_{Co_{1}}$ and $z_{Co_{2}}$, the distance between the cobalt atoms $d_{Co-Co}$, the angle $\theta$ between the dimer axis and the norm of the circumcoronene plane, the magnetic moment of cobalt atoms separately $\mu_{Co_{1}}$ and $\mu_{Co_{2}}$.
For Co$_{4}$ clusters it is shown the distance between each cobalt atom and its first carbon neighbour $d_{Co-C}$ together with the magnetic moment of that cobalt atom.
}
\end{table*}

We start reviewing the magnetic properties focusing first on the free systems. For the circumcoronene molecule we obtained a non-magnetic equilibrium state, with no localized magnetic moment on edge atoms, which agrees with results of previous work in base to the size of circumcoronene \cite{fernandez2007magnetism}. Note that it is an aromatic molecule following Clar's rule \cite{popov2012chemical}.
For the free cobalt atom the magnetic moment in the equilibrium state is 3 $\mu_{B}$. When we allowed charge division, we obtain that an electron configuration like $[Ar] 3d^{7.67} 4s^{1.33}$ stabilizes the system. Based on these results we generate the pseudopotential that we are using here in SIESTA calculations.
Concerning the cobalt dimer we obtained a bond length of approximately 2 \AA{}, with a total magnetic moment of 4 $\mu_{B}$, as in Ref \cite{xiao2009co}. However, SIESTA calculations obtain values about than 5 $\mu_{B}$.
The bonding energy E$_{B}$ for the cobalt molecule is 1.757 eV.
For the Co$_{4}$ cluster, the magnetic moment is 10 $\mu_{B}$. The structure is called butterfly wings (bw) after its shape. Almost square in shape, the bond lengths are 2.208 \AA{} and 2.859 \AA{}. SIESTA drops values quite similar, of 2.19 \AA{} and 2.84 \AA{}. The values of the magnetic moments on each atom are the same at geometrically equivalent sites because there is no charge transfer found between cobalt atoms.
The bonding energy E$_{B}$ for the four cobalt atoms is 2.440 eV.

In order to understand why we do not get the same result as with ADF, we have checked a large amount of radii for the pseudopotential generation without getting closer to the 4 $\mu_{B}$ value for Co$_2$ molecule. Even when we reduced the radii as much as possible to have a hard but highly transferable pseudopotential, we obtain the desired value of magnetization, but we spoiled the good results for larger systems, as Co$_{4}$ clusters or cobalt bulk.
We check that choosing the pseudopotential relativistic or not is important in this particular case. We obtain that for non-relativistic pseudopotentials the order of the levels corresponds to a 4 $\mu_{B}$ state, changing the magnetization because a very small broadening would be required to get closer to the 4 $\mu_{B}$ value.
Then, we create a pseudopotential with 3p electrons included as valence electrons, but we did not yet get the desired 4 $\mu_{B}$ value. Finally, by including the 3s electrons we got the reported as correct magnetic moment value \cite{xiao2009co}. To get this value of the magnetic moment the FPLO code requires to include the 3s electrons as valence electrons.
However, we note that another option to get closer to the 4 $\mu_{B}$ value is to change the electric configuration from 4s$^{1.5}$3d$^{7.5}$ to 4s$^{1}$3d$^{8}$. In this way we obtain values slightly larger than 4 $\mu_{B}$, which also improves previous results.
From all this discussion, the conclusion is that the cobalt dimer stands forward as critical to be modeled using pseudopotentials. To obtain the magnetic moment of 4 $\mu_{B}$ requires reducing the number of core electrons, which in fact is moving towards a full-electron calculation, as it is the case of using ADF or FLPO codes \cite{xiao2009co}.
Anyhow, for larger systems, pseudopotentials are not so crucial, and similar results are obtained with different optimized pseudopotentials. In this work we are using the pseudopotential we obtained that provided the smallest energy differences among several electronic configurations during the tests we carried out. Including 3s and 3p electrons would improve our results for larger systems, but only slightly; this improvement does not justify the loss of computation speed, taking into account that we chose SIESTA on purpose to study in detail many adsorption geometries and to discuss general trends, more than focusing on a specific configuration.
Nevertheless, this issue with Co$_{2}$ molecule shows useful to know the limitations of our choosen pseudopotential.

We next proceed to deposit Co atoms, Co molecule and Co$_4$ clusters around the center of circumcoronene.
The bonding energy is computed as $E_{b}=(E_{C_{54}H_{18}}+E_{Co}-E_{C_{54}H_{18}+Co_{n}})/n$, with n=1,2,4.
In the central ring of circumcoronene, Co atom is at the central hollow site (CH) with a height of 1.5 \AA{}, establishing six bonds to the carbon atoms with a bond length of 2.13 \AA{}. The magnetic moment of the system amounts to 1 $\mu_{B}$. The SIESTA results for this configuration present similar values for bond heights and lengths of 2.105 \AA{} and 1.454 \AA{} respectively.

For Co$_{2}$ on circumcoronene we tried three configurations in top, bridge, and hollow sites (lCT, lCB and pCH). The most stable is the perpendicular pCH hollow configuration, with a cobalt atom placed at the centre of the ring, with a bond length of 2.28 \AA{}, slightly larger than in SIESTA results where it was 2.201 \AA{}. The height over circumcoronene is 1.77 \AA{}, comparable to 1.668 \AA{} obtained with SIESTA. However, the cobalt-cobalt length becomes 2.37 \AA{} obtained with ADF shorter from  2.12 \AA{} with SIESTA. These trends propagate the magnetic moment of Co molecule once adsorbed on circumcoronene. For the ADF calculation, the magnetic moment is zero because the cobalt atoms are anti-ferromagnetically coupled, while it becomes ferromagnetic with a value of 4 $\mu_{B}$ in the SIESTA calculation. Table~\ref{tbl:adf} gives the adsorption energy of the pCH configuration in ADF code about 0.042 eV, which indicates that the bonding process could be considered to be physisorption more than chemisorption, as in the rest of cases. Stretching slightly the distance Co-Co in SIESTA calculations we end in the same zero antiferromagnetic values; this indicate that in the search in previous ADF calculations of Co molecules on circumcoronene, we have overlooked the minimum energy for the ferromagnetic configuration.  

We obtained that placing the cobalt dimer perpendicularly over the central hollow site (pCH) results in a more stable geometry than placing it lying on the bridge (lCB) or top (lCT) sites around the central hollow site. In fact the adsorption energies for the lCB and lCT configurations are negative, which indicates that not only the pCH configuration is more stable but that is preferred over the free atoms. The same stability order is reproduced with the SIESTA calculations in Table III for those three configurations.

We last compare ADF geometries with those obtained using SIESTA. In the case of lCB configuration, the cobalt height over circumcoronene differs by about 0.1 \AA{}, thus the cobalt-carbon bond length is slightly larger.
Anyhow, the values of the magnetic moment are quite similar within 0.1 $\mu_B$. 
For the lCT configuration, the cobalt height and the cobalt-carbon bond length are shorter, while the cobalt-cobalt bond length is slightly larger.
The magnetic moment values are also similar, but in ADF results the magnetic moment of cobalt atoms and the induced magnetic moment on circumcoronene are smaller.
For both lCT and lCB configurations, the total magnetic moment in ADF calculations is 4 $\mu_{B}$, which is the same to 4.0 $\mu_{B}$ obtained in SIESTA calculations.

We compute Co$_{4}$ clusters in the circumcoronene center. We tried three configurations starting from the found ground isomer of Co$_{4}$ clusters. Placing the Co cluster around the central hollow site,
the bw structure lies (i) with an edge parallel to circumcoronene and its two cobalt body atoms over bridge sites in the bwCB case or (ii) on top sites in the bwCT one, or (iii) joined by a wing face with the three cobalt atoms on alternate bridges. The third case becomes the most stable after structure relaxations. The outer cobalt atom moved towards the side center and the hollow site to build a regular-like tetrahedron. This outer atom ends with a cobalt-carbon distance of 2.11 \AA{}, similar to the one obtained in SIESTA (2.05 \AA{}), and the magnetic moment is 6 $\mu_{B}$, identical in both codes.

For the bwCB and bwCT configurations the values of the magnetic moment in ADF and SIESTA are also similar, but now two pairs of cobalt atoms behave symmetrically in the same way. In all the cases the charge transfer goes from cobalt atoms to carbon atoms.
The bwCB and bwCT configurations have a total magnetic moment of 8 $\mu_{B}$, larger than for atoms in the CB configuration. The adsorption energy for the bwCB case is almost zero and negative for bwCT. This means that the cobalt Co$_4$ cluster prefers to be in the free-state rather than to be adsorbed in those unstable configurations.



\end{document}